\documentclass[journal,12pt,onecolumn,draftclsnofoot]{IEEEtran}
\usepackage{amsmath,amsfonts}
\usepackage{algorithmic}
\usepackage{algorithm}
\usepackage{array}
\usepackage[caption=false,font=normalsize,labelfont=sf,textfont=sf]{subfig}
\usepackage{textcomp}
\usepackage{stfloats}
\usepackage{url}
\usepackage{verbatim}
\usepackage{graphicx}
\usepackage{cite}
\usepackage{xcolor}
\usepackage{tabularx}
\newtheorem{lem}{\bf Lemma}

\newtheorem{prob}{\bf Problem}

\begin{document}

\title{Joint Task Offloading and Resource Allocation for Streaming Application in Cooperative Mobile Edge Computing}

\author{Xiang Li, Rongfei Fan, {\em Member, IEEE}, \\ Han Hu, {\em Member, IEEE}, and Xiangming Li, {\em Member, IEEE}

\thanks{
	Corresponding author: Rongfei Fan (fanrongfei@bit.edu.cn).
	
	Xiang Li and Han Hu are with the School of Information and Electronics, Beijing Institute of Technology, Beijing 100081, P. R. China. (\{lawrence,hhu\}@bit.edu.cn).
	
	Rongfei Fan and Xiangming Li is with the School of Cyberspace Science and Technology, Beijing Institute of Technology, Beijing, 10081, P. R. China. ({fanrongfei,xmli}@bit.edu.cn).
	
	
}
}

\markboth{Journal of \LaTeX\ Class Files,~Vol.~14, No.~8, August~2021}%
{Shell \MakeLowercase{\textit{et al.}}: A Sample Article Using IEEEtran.cls for IEEE Journals}

\IEEEpubid{0000--0000/00\$00.00~\copyright~2021 IEEE}

\maketitle

\begin{abstract}
Mobile edge computing (MEC) enables resource-limited IoT devices to complete computation-intensive or delay-sensitive task by offloading the task to adjacent edge server deployed at the base station (BS), thus becoming an important technology in 5G and beyond.
Due to channel occlusion, some users may not be able to access the computation capability directly from the BS.
Confronted with this issue, many other devices in the MEC system can serve as cooperative nodes to collect the tasks of these users and further forward them to the BS.
In this paper, we study a MEC system in which multiple users continuously generate the tasks and offload the tasks to the BS through a cooperative node.
As the tasks are continuously generated, users should simultaneously execute the task generation in the current time frame and the task offloading of the last time frame, i.e. the task is processed in a streaming model.
To optimize the power consumption of the users and the cooperative node for finishing these streaming tasks, we investigate the duration of each step in finishing the tasks together with multiuser offloading ratio and bandwidth allocation within two cases: the BS has abundant computation capacity (Case I) and the BS has limited computation capacity (Case II).
For both cases, the formulated optimization problems are nonconvex due to fractional structure of the objective function and complicated variable coupling.
To overcome this challenge, in Case I, the objective function is transformed with Dinkelbach method.
Then the multiuser offloading ratio and bandwidth allocation, the duration of each step are optimized separately, for which bisection search and interior-point method are utilized to reach the optimal solution.
In case II, the problem is even more complex due to the constraint of the BS's computation capacity.
Regarding this issue, convergent solution is searched levaraging difference of convex algorithm.
Finally, simulation is carried out to verify the effectiveness of the proposed methods and reveal the performance of the considered system.

\end{abstract}


\section{Introduction}
\IEEEPARstart{W}{}ith the rapid development of deep learning technology, recent years have witnessed a drastically booming of smart mobile applications.
In order to alleviate the pressure posed by these computation-intensive and delay-sensitive mobile applications, mobile edge computing (MEC) is proposed by European Telecommunications Standards Institute (ETSI) and has been  widely recognized as a promising technology in 5G and beyond.
By offloading the computation task to the nearby computation server implemented in the base station or the network access point, not only can the application be executed within a rather low latency, but the energy consumption of the mobile device can also be reduced significantly.

On the other hand, many of these smart mobile applications rely on the technology of video analytics.
Video analytics is a process of analyzing video content to identify and detect different objects in video streams, it has been used in a wide range of applications such as facial recognition, behavior detection, and traffic monitoring.
In terms of the working principle of video analytics, generally they are performed by sampling the video at a certain interval and then analyzing the pictures generated by the sampling.
According to the working principle, the video analytic applications have several characteristics:
First, the pictures obtained by sampling is still large in data size and the processing of these pictures is dependent upon deep neural networks, which generally implies great computation complexity.
Second, the specific task of these applications is formulated with the data collected over a time period, and the overall collection of the data may span several period.
Take traffic monitoring as an example, the surveillance equipment will continuously monitor the road, but execute an analytic task every few moments.
Since these applications are mostly executed on wireless terminal such as cell phone and smart monitoring device, which is limited either in energy supply or in computation capacity,
based on these characteristics, it is a natural choice to utilize MEC for the application of video analytics.

Despite the potential benefits of leveraging MEC for video analytics in terms of conserving energy and computing resources of mobile devices, existing research on MEC has not adequately addressed the need to re-model computation offloading and completion to account for the specific characteristics of these applications.
Most research on MEC only considers the completion of a given task with fixed data size and computation complexity while ignoring the process of data generation.
In practice, tasks should be formulated by collecting the data accumulated within a time frame in which local computing can be executed.
In the following time frame, the task can be offloaded to the base station while data of the next task is being accumulated. 
This execution process follows a streaming task model, carried out in an assembly line fashion similar to the execution of streaming algorithms. 
In this streaming model, contrary to the widely researched given task completion model, the data accumulating time frame significantly influences the system performance such as energy consumption or time latency for finishing the task. 

This paper is intended for optimizing the power consumption for the execution of streaming tasks where the task is continuously generated.
Specifically, multiple mobile users collect data to formulated their computation tasks and offload part of the task to the BS through a cooperative node, which is responsible for collecting the channel state and task information and adjusting the resource allocation.
As the cooperative node is mostly a laptop or other wireless equipment that has limited battery capacity, the overall cost function to be optimized concerns the power consumption of both the users and the cooperative node.
In order to achieve lower power consumption, the time length of collecting the data, transmission from the user to the cooperative node, transmission from the cooperative node to the BS, and the remote execution on the BS are jointly adjusted, together with the offloading ratio and bandwidth allocation for each user.

The contributions of this paper are summarized as follows:
\begin{itemize}
	\item We study the streaming model to account for the execution of continuously generated computation tasks in MEC for the first time. Regarding this streaming model, we propose a cooperative node assisted MEC system for multiple users each of which has task to finish.
	\item We consider the power consumption minimization in two cases, i.e. when the BS has abundant computation capacity (Case I) and when the BS has limited computation capacity (Case II).
	The optimization problem formulated in both cases are fractional programming problem that are generally difficult to solve.
	To this end, we transform the problem into a tractable form with a Dinkelbach-based method.
	Afterwards, in case I, we utilize block coordinate descent (BCD) method to optimize the duration of each step and the resource allocation of multiple users separately, then analyze the solution of multiuser resource allocation with lagrangian multiplier method. Through this method, the complexity of finding the optimal solution can be reduced remarkably compared with conventional method.
	In case II, the problem is even more complicated than Case I since the constraint of computation capacity exacerbate the coupling between variables. To this end, we apply difference of convex algorithm (DCA) to address the convergent solution by iteration. Through this method, local optimum of the problem can be found by a proper complexity.
	\item We verify the effectiveness and convergence of the proposed algorithm for Case I and Case II with simulation. Finally, to evaluate the performance of the proposed system, the optimal offloading ratio and bandwidth allocation of the users, together with the average power consumption of the users and the cooperative node, are depicted under the setting of different system parameters.
\end{itemize}

The rest of the paper is organized as follows.
Section \ref{s:related_work} reviews related work.
Section \ref{s:system_model} introduces the system model and formulate the optimization problems for Case I and Case II.
Section \ref{s:optimal_solution_first} presents the problem transformation and solution related to Case I.
Section \ref{s:optimal_solution_second} demonstrates the problem solution corresponding to Case II.
Section \ref{s:numerical_results} illustrates the numerical results.
Finally, Section \ref{s:conclusion} summarizes the conclusion remarks.

\section{Related Work} \label{s:related_work}
Mobile edge computing, energy consumption and execution latency minimization.

Little research has been focused on the streaming process of the task computation and the task offloading.
Most streaming only talk about video streaming.
Cooperative node assisted are even less.

\section{System Model and Problem Formulation} \label{s:system_model}
\subsection{Multiuser task offloading model}
\begin{figure}
	\begin{center}
		\includegraphics[angle=0,width=0.47 \textwidth]{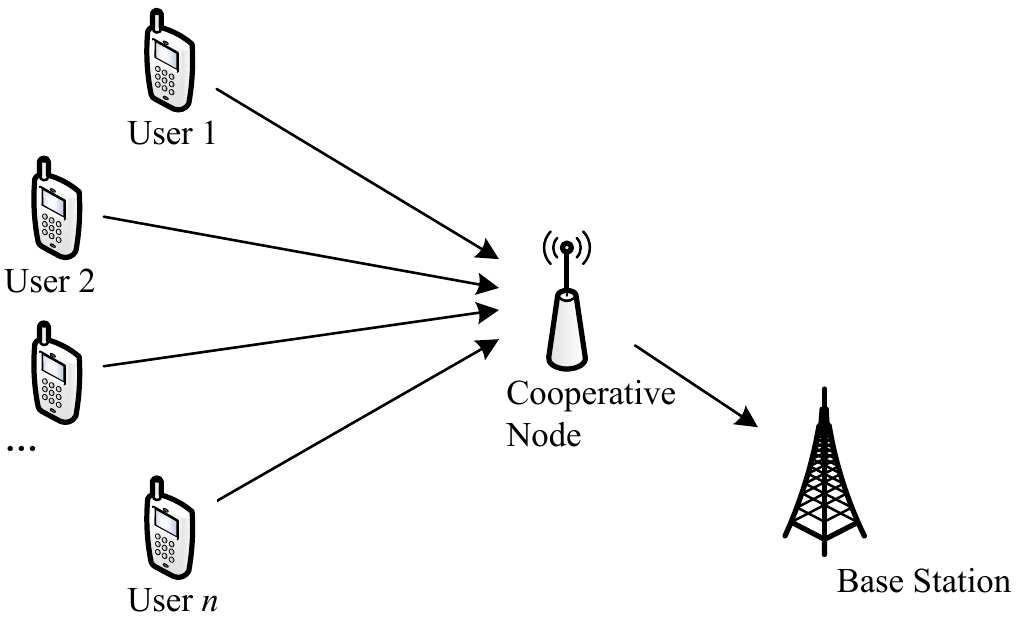}
	\end{center}
	\caption{Task offloading of the IoT device and the cooperative node.}
	\label{f:all_structure}
\end{figure}
Consider a multi-user MEC system where each user collects data and generates the computation tasks with the collected data.
There is a cooperative node that serves as a task data collector of these multiple users and further forwards these tasks to the BS, which is mounted with an edge server.
The computation task of the users can be executed in two ways:
1) The users execute part of the task in real time while collecting the data. Hereinafter, this part will be referred to as local computation.
2) The users offload part of the task to the BS through the cooperative node, and the BS finishes the computation. Hereinafter, this part will be referred to as edge computation.
The system is shown in Fig. \ref{f:all_structure}.

Since the users' data accumulate over time and the task is generated with the accumulated data, 
the data size of the computation task is proportional to the accumulating time length, which is denoted by $t_1$.
After $t_1$, the users send their task data to the cooperative node with time $t_2$ and the cooperative node, upon receiving the data, sends it to the BS with time $t_3$.
The role of the cooperative node are two-folds:
\begin{itemize}
	\item For most smart applications, the content service provider and the infrastructure service provider are usually different.
	The cooperative node is local device and the BS (and the edge server) is deployed by the infrastructure service provider. 
	To make the task accessible for the devices other than the content service provider, the cooperative node is responsible for the pre-processing of the computation task before the task is transferred to a third-party computation resource. 
	\item 
	The cooperative node can collect the information of the computation tasks and the channel between the users, the cooperative node and the BS, in order to provide an efficient way to finish the task.
	Furthermore, the channel state between the users and the BS is unreliable sometimes, and the cooperative node can serve as a relay to enhance the transmission.
\end{itemize}
Finally, the BS receives the task and finishes the computation within time $t_4$.
In this work, we consider two different cases of the BS 's computation capacity: 1) The BS has abundant computation capacity. This case is applicable when the computation required to finish the task is much smaller than the computation capability of the BS. In this case, $t_4$ can be recognized as 0. 
2) The BS has finite computation capacity. This case is related to the situation when the computation required to finish the task is comparable to the computation capability of the BS. In this case, allocation of $t_4$ should be considered in the system.
Since the computation result is usually quite small in data size, the delay for sending back the result to the users is ignored \cite{Self_dependency}.

\begin{figure}
	\begin{center}
		\includegraphics[angle=0,width=0.47 \textwidth]{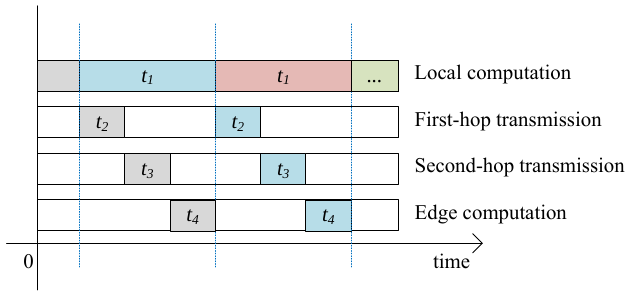}
	\end{center}
	\caption{Task offloading of the IoT device and the cooperative node.}
	\label{f:time_order}
\end{figure}
Since the users process streaming task, the data of the task is continuously generated.
As shown in Fig. \ref{f:time_order}, the users formulate and send their tasks to the cooperative node at regular intervals $t_1$.
To simplify the expression, the duration of $t_1$ is called a slot.
The users execute local computation while collecting the data in the current slot, and then offload the rest of the tasks to the BS to finish them with the help of the cooperative node in the next slot.
In order to formulate a task, the minimum time to collect the data is $t^l$.
Since the users have limited storage capacity, the slot length should be no larger than $t^u$.
Thus, there is
\begin{equation}
	t^l \leq t_1 \leq t^u
\end{equation}
To ensure that the task data is not backlogged either at local or at the BS, the remaining part of task generated in the current slot should be finished within the next slot by the BS.
When the BS has finite computation capacity, there is 
\begin{equation}
t_2 + t_3 + t_4 \leq t_1
\end{equation}
On the contrary, when the BS has abundant computation capacity, there is $t_4=0$. Thus
\begin{equation}
t_2 + t_3 \leq t_1
\end{equation}

\subsection{Computing and transmitting model}
Define the set of users as $\mathcal{N} = \{1, 2, \cdots, N\}$.
For user $n$, 
the data amount of its task accumulated within unit time second is $c_n$, i.e. the data size of the task is $c_n t_1$, which is in the unit of nat. 
The computation capacity to finish a nat of task is denoted as $X$, which is expressed by the number of required CPU cycles.
Similar as in \cite{Self_multi_relay, Cui_wpt}, 
for a general CPU working at frequency $f$, which means that it can process $f$ CPU cycles per second, the power consumption can be written as $\kappa f^3$, where $\kappa$ is the computation energy efficiency coefficient of the processor's chip.
Suppose user $n$ has an offloading ratio of $r_n$, i.e. it offloads $r_n$ of the task to the BS, and compute $(1-r_n)$ of the task locally.
To finish the local computation on user $n$, the required computation capacity is $t_1 X c_n (1-r_n)$, and the relative time latency is $t_1$, thus the CPU frequency is $X c_n (1-r_n)$.
Then the energy consumption $E^l_n$ of the local computation at user $n$ is
\begin{equation} \label{e:user_comp_e}
E^l_n = \kappa_n t_1 X^3 {c_n}^3 {(1-r_n)}^3 + E^o_n, \forall n \in \mathcal{N}
\end{equation}
in which $\kappa_n$ is the energy coefficient of user $n$.
Besides, the overhead $E^o_n$ includes the energy consumption by all other components such as memory collecting, which is a parameter regardless of the speed of the processor.

On the other hand, in order to offload the task to the BS, the users and the cooperative node should transmit in a sequential order.
First, the users send their data to the cooperative node with frequency-division multiple access (FDMA) in a duration of $t_2$.
Suppose the transmission of user $n$ takes up a bandwidth of $b_n$ and the system bandwidth is $B$, there is 
\begin{equation}
\sum_{n} b_n \leq B
\end{equation}
Assume the channel gain between user $n$ and the cooperative node on subband $n$ is $h_n$ 
\footnote{In this paper, the concept `channel gain' denotes the uniformed signal-to-noise ratio, which is actually the ratio of the channel gain to the power spectrum density of noise.}, 
and the transmit energy consumption of user $n$ is $E^t_n$.
Applying Shannon's formula, there is
\begin{equation} \label{e:user_tran_d}
t_1 c_n r_n =
t_2 b_n \ln \left(1+\frac{h_n E^t_n}{t_2 b_n}\right), \forall n \in \mathcal{N}
\end{equation}
which indicates that
\begin{equation} \label{e:user_tran_e}
E^t_n = \frac{t_2 b_n}{h_n} \left( e^{\frac{t_1 c_n r_n}{t_2 b_n}} -1 \right), \forall n \in \mathcal{N}
\end{equation}

After receiving the information from users, the cooperative node send the collected task data to the BS, which occupies the overall system bandwidth $B$ and lasts for a duration of $t_3$.
Suppose the channel gain between the cooperative node and the BS is $g$, and the transmit energy consumption of the cooperative node is $E^t_c$. 
Following similar discussion as with (\ref{e:user_tran_e}), there is
\begin{equation} \label{e:coop_tran_e}
E^t_c = \frac{t_3 B}{g} \left( e^{\frac{t_1 \sum_{n} c_n r_n}{t_3 B}} -1 \right)
\end{equation}

Upon receiving the tasks from the cooperative node, the BS needs to complete the computation of all the users' tasks within time $t_4$. 
When the BS has finite computation capacity, suppose the computation capacity allocated by the BS to finish these tasks is $f_B$. Then there should be
\begin{equation} \label{e:bs_computation}
\sum_{n = 1}^N \frac{t_1 X c_n r_n}{t_4} \leq f_B
\end{equation}
On the other hand, when the BS has abundant computation capacity, the task will be finished within an extremely short delay that is negligible.
Thus this constraint in (\ref{e:bs_computation}) is no longer needed.

\subsection{Problem formulation}
In this system, the goal of our research is to achieve the most energy-efficient design of both the users and the cooperative node.
Since the data of the computation task is generated continuously, i.e. it is in a steaming structure,
the specific size of the task is related to the value of $t_1$.
Thus it is no longer meaningful to simply investigate the energy consumption of completing a specific task.
Instead, we need to work on the ratio of energy consumption to the time for completing the task, which will be referred to as average power consumption in the following context.
Therefore, the objective function in this work is the ratio of the overall energy consumption of the users and the cooperative node to the time latency for finishing the tasks.
The users' energy consumption consists of two parts: one is the cost of local computation as in (\ref{e:user_comp_e}), the other one is the energy required for offloading the tasks to the cooperative node as in (\ref{e:user_tran_e}).
The cooperative node's energy is consumed in offloading the users' tasks from the cooperative node to the BS, as in (\ref{e:coop_tran_e}).
In terms of the latency for finishing the tasks: 
when the BS has finite computation capacity, the latency is $t_1+t_2+t_3+t_4$, 
when the BS has abundant computation capacity, the latency is $t_1+t_2+t_3$.

In order to minimize the objective function, we consider the evaluation of $t_1$, $t_2$, $t_3$ (and when the BS has finite computation capacity, $t_4$), together with the offloading ratio and bandwidth allocation of the users $\{r_n, b_n|n \in \mathcal{N}\}$.
When the BS has abundant computation capacity, the optimization problem can be formulated as follows:
\begin{prob} \label{p:first}
	\begin{subequations}
		\begin{align}
			\min_{\substack{t_1, t_2, t_3, \\ \{r_n, b_n|n \in \mathcal{N}\}}}
			& \frac{\sum_{n} \frac{t_2 b_n}{h_n} \left( e^{\frac{t_1 c_n r_n}{t_2 b_n}} -1 \right) + \frac{t_3 B}{g} \left( e^{\frac{t_1 \sum_{n} c_n r_n}{t_3 B}} -1 \right) +\sum_{n} \left( \kappa_n t_1 X^3 {c_n}^3 {(1-r_n)}^3 +E^o_n \right)}{t_1 +t_2 +t_3} \notag \\
			\text{s.t.} \quad
			& t^l \leq t_1 \leq t^u, \label{e:first_con_time_lower_upper} \\
			& t_2 +t_3 \leq t_1, \label{e:first_con_time} \\
			& \sum_{n = 1}^N b_n \leq B,  \label{e:first_con_band} \\
			& 0 \leq r_n \leq 1, n \in \mathcal{N}, \label{e:first_con_r}\\
			& b_n \geq 0, n \in \mathcal{N}. \label{e:first_con_b}
		\end{align}
	\end{subequations}
\end{prob}
In Problem \ref{p:first},
(\ref{e:first_con_r}) is the box constraint with respect to the offloading ratio, and (\ref{e:first_con_b}) is the positive constraint of bandwidth allocated to the users.
The difficulty in solving Problem \ref{p:first} lies in several aspects:
1) The objective function is in a fractional structure, which is obviously nonconvex. 
2) There are terms including coupling between optimization variables $\{t_1, t_2, t_3\}$ and $\{r_n\}$, $\{b_n\}$ in the objective function.
3) The sum of $\{r_n\}$ appearing on the exponent in the objective function make the problem unable to be solved separately with different $n$. 

When the BS has finite computation capacity, similar as the abundant-capacity counterpart, the optimization problem can be formulated as follows:
\begin{prob} \label{p:second}
	\begin{subequations}
		\begin{align}
			\min_{\substack{t_1, t_2, t_3, t_4, \\ \{r_n, b_n|n \in \mathcal{N}\}}}
			& \frac{\sum_{n} \frac{t_2 b_n}{h_n} \left( e^{\frac{t_1 c_n r_n}{t_2 b_n}} -1 \right) + \frac{t_3 B}{g} \left( e^{\frac{t_1 \sum_{n} c_n r_n}{t_3 B}} -1 \right) +\sum_{n} \left( \kappa_n t_1 X^3 {c_n}^3 {(1-r_n)}^3 +E^o_n \right)}{t_1 +t_2 +t_3 +t_4} \notag \\
			\text{s.t.} \quad
			& t^l \leq t_1 \leq t^u, \label{e:second_con_time_lower_upper} \\
			& t_2 +t_3 +t_4 \leq t_1, \label{e:second_con_time} \\
			& \sum_{n = 1}^N t_1 X c_n r_n \leq f_B t_4, \label{e:second_con_comp} \\
			& \sum_{n = 1}^N b_n \leq B,  \label{e:second_con_band} \\
			& 0 \leq r_n \leq 1, n \in \mathcal{N}, \label{e:second_con_r}\\
			& b_n \geq 0, n \in \mathcal{N}. \label{e:second_con_b}
		\end{align}
	\end{subequations}
\end{prob}
In Problem \ref{p:second}, constraint (\ref{e:second_con_comp}) is transformed from (\ref{e:bs_computation}).
Asides from the difficulty lying as in Problem \ref{p:first}, in Problem \ref{p:second}, variable $t_1$, $\{r_n\}$ and $t_4$ are also coupled in constraint (\ref{e:second_con_comp}), which make the Problem more complex to solve.

In the following context, for the purpose of simplification, the case when the BS has abundant computation capacity and finite computation capacity will be referred to as Case I and Case II, respectively.
Within the next two sections, we will discuss the solution corresponding to these two cases.

\section{Solution for the case when the BS has abundant computation capacity} \label{s:optimal_solution_first}

In this section, we will address the difficulties of Problem \ref{p:first} and find the local optimal solution at low computation complexity.

\subsection{Problem transformation} \label{s:first_problem_transformation}
To dual with the fractional structure in the objective function of Problem \ref{p:first}, the most straightforward approach is to find a way to decouple the numerator and the denominator and transform the problem into a simpler case.
Thus, the objective function of Problem \ref{p:first} can be reformulated by utilizing a similar procedure as Dinkelbach method.

\sloppy
To simplify the expression, denote the optimization variables of Problem \ref{p:first}, i.e. $t_1, t_2, t_3, \{r_n, b_n|n \in \mathcal{N}\}$, as $\boldsymbol{x}$. Define the numerator and the denominator of Problem \ref{p:first} 's objective function as ${N}(\boldsymbol{x})$ and ${D}(\boldsymbol{x})$, respectively.
By introducing a slack variable $\theta$ such that $\frac{{N}(\boldsymbol{x})}{{D}(\boldsymbol{x})} \leq \theta$, Problem \ref{p:first} can be equivalently transformed into:
 \begin{prob} \label{p:dinkelbach_proof}
	\begin{subequations}
		\begin{align}
		\min_{\theta, \boldsymbol{x}} \quad
		& \theta \notag \\
		\text{s.t.} \quad
		& {N}(\boldsymbol{x}) - \theta {D}(\boldsymbol{x}) \leq 0 \label{e:dinkelbach_proof_con} \\
		& (\ref{e:first_con_time_lower_upper})-(\ref{e:first_con_b}).
		\end{align}
	\end{subequations}
\end{prob}
For Problem \ref{p:dinkelbach_proof}, by denoting the lagrangian variable related to (\ref{e:dinkelbach_proof_con}) as $\xi$ and applying Karush-Kuhn-Tucker(KKT) conditions, the optimal solution in the feasible region of $(\ref{e:first_con_time})-(\ref{e:first_con_b})$ should satisfies the following equations:
\begin{subequations}
\begin{align}
1- \xi {D}(\boldsymbol{x}) = 0 \\
\xi \left( {N}(\boldsymbol{x}) - \theta {D}(\boldsymbol{x}) \right) = 0
\end{align}
\end{subequations}
Since ${D}(\boldsymbol{x}) \neq 0$ and the lagrangian variable $\xi \geq 0$, it can be established that the optimal solution of $\boldsymbol{x}$ and multiplier $\xi$ satisfies ${N}(\boldsymbol{x}) - \theta {D}(\boldsymbol{x}) = 0$.
Accordingly, with an initial value of $\theta$, Problem \ref{p:dinkelbach_proof} can be solved by alternatively executing the two steps below \footnote{Similar discussion and solution can be also found in \cite{vm,latency_ris_mec}}.
	
1) Solve Problem \ref{p:first_minus_simplify} to get optimal $\boldsymbol{x}$:
\begin{prob} \label{p:first_minus_simplify}
	\begin{subequations}
		\begin{align}
		\min_{\boldsymbol{x}} \quad
		& {N}(\boldsymbol{x}) - \theta {D}(\boldsymbol{x}) \notag \\
		\text{s.t.} \quad
		& (\ref{e:first_con_time_lower_upper})-(\ref{e:first_con_b}).
		\end{align}
	\end{subequations}
\end{prob}

2) Let $\theta = \frac{{N}(\boldsymbol{x})}{{D}(\boldsymbol{x})}$.

In the above two steps, the second step is quite straightforward. We only focus on the first step, i.e. solution of Problem \ref{p:first_minus_simplify}, whose complete form is
\begin{prob} \label{p:first_minus}
	\begin{subequations}
		\begin{align}
		\min_{\substack{t_1, t_2, t_3, \\ \{r_n, b_n|n \in \mathcal{N}\}}}
		& {\sum_{n} \frac{t_2 b_n}{h_n} \left( e^{\frac{t_1 c_n r_n}{t_2 b_n}} -1 \right) + \frac{t_3 B}{g} \left( e^{\frac{t_1 \sum_{n} c_n r_n}{t_3 B}} -1 \right) +\sum_{n} \kappa_n t_1 X^3 {c_n}^3 {(1-r_n)}^3} \notag \\
		& + \sum_{n} E^o_n - \theta {\left(t_1 +t_2 +t_3\right)} \notag \\
		\text{s.t.} \quad
		& (\ref{e:first_con_time_lower_upper})-(\ref{e:first_con_b}).
		\end{align}
	\end{subequations}
\end{prob}
In Problem \ref{p:first_minus}, the objective function is nonconvex since variable $t_1, t_2$ and $\{r_n, b_n|n \in \mathcal{N}\}$ are coupled, which makes the problem unable to be solved globally. In this respect, we utilize block coordinate descent (BCD) method to find the local optimal solution by convergence.
Specifically, the solution of Problem \ref{p:first_minus} can be obtained with an iterative procedure.
At each iteration, two sets of variables are optimized in consecutive order:
\begin{itemize}
	\item In the first step, the variables $\{r_n, b_n|n \in \mathcal{N}\}$ are optimized with fixed $t_1, t_2, t_3$, which is referred to as offloading ratio and bandwidth optimization. This part will be discussed in Subsection \ref{s:first_problem_bandr}.
	\item In the second step, the variables $t_1, t_2, t_3$ are optimized with fixed $\{r_n, b_n|n \in \mathcal{N}\}$, which is referred to as optimal time allocation. This part will be discussed in Subsection \ref{s:first_problem_t}.
\end{itemize}

\subsection{Offloading ratio and bandwidth optimization} \label{s:first_problem_bandr}
In this subsection, we will introduce the solution of $\{r_n, b_n|n \in \mathcal{N}\}$. 
With fixed $t_1, t_2, t_3$ that satisfies (\ref{e:first_con_time_lower_upper}) and (\ref{e:first_con_time}), it can be checked that the objective function of Problem \ref{p:first_minus} is a jointly convex function of $\{r_n, b_n|n \in \mathcal{N}\}$,
and the related constraints (\ref{e:first_con_band})-(\ref{e:first_con_b}) are linear with $\{r_n, b_n|n \in \mathcal{N}\}$.
Therefore, Problem \ref{p:first_minus} becomes a convex optimization problem which can be solved by existing numerical methods, e.g. interior-point method \cite{numerical_book}.
To bring more analytical insights and reduce computation complexity, we utilize lagrangian method to analyze and discuss the optimal structure of the solution.

Before going into details of the optimal structure, to simplify the solution, we introduce auxiliary variable $z$ such that $\sum_{n} c_n r_n \leq z$.
Then Problem \ref{p:first_minus} can be equivalently transformed into Problem \ref{p:b_and_r}.
\begin{prob} \label{p:b_and_r}
	\begin{subequations}
		\begin{align}
		\min_{\substack{z, \{r_n, b_n|n \in \mathcal{N}\}}}
		& {\sum_{n} \frac{t_2 b_n}{h_n} \left( e^{\frac{t_1 c_n r_n}{t_2 b_n}} -1 \right) + \frac{t_3 B}{g} \left( e^{\frac{t_1 z}{t_3 B}} -1 \right) +\sum_{n} \kappa_n t_1 X^3 {c_n}^3 {(1-r_n)}^3} \notag \\
		& + \sum_{n} E^o_n - \theta {\left(t_1 +t_2 +t_3\right)} \notag \\
		\text{s.t.} \quad 
		& \sum_{n} c_n r_n \leq z \label{e:b_and_r_con_z} \\
		& (\ref{e:first_con_band})-(\ref{e:first_con_b}).
		\end{align}
	\end{subequations}
\end{prob}
The equivalent relationship between Problem \ref{p:first_minus} and Problem \ref{p:b_and_r} can be recognized this way:
The optimal solution of $z$ in Problem \ref{p:b_and_r} should satisfy $\sum_{n} c_n r_n = z$. If not, one can always achieve lower objective function value by replacing $z$ with $\sum_{n} c_n r_n$, which is certainly feasible when $z$ is in the feasible region of Problem \ref{p:b_and_r}.

Then we analyze Problem \ref{p:b_and_r} with lagrangian method.
Denote the lagrangian multiplier related to constraint (\ref{e:b_and_r_con_z}) and (\ref{e:first_con_band}) as $\varepsilon$ and $\lambda$, respectively, the partial Lagrangian of Problem \ref{p:b_and_r} can be derived as:
\begin{align}
\mathcal{L} = & {\sum_{n} \frac{t_2 b_n}{h_n} \left( e^{\frac{t_1 c_n r_n}{t_2 b_n}} -1 \right) + \frac{t_3 B}{g} \left( e^{\frac{t_1 z}{t_3 B}} -1 \right) +\sum_{n} \kappa_n t_1 X^3 {c_n}^3 {(1-r_n)}^3} \\
& + \sum_{n} E^o_n - \theta {\left(t_1 +t_2 +t_3\right)} + \lambda \left( \sum_{n} b_n -B \right) + \varepsilon \left( \sum_{n} c_n r_n -z \right)
\end{align}
With this expression of Lagrangian, the associated dual function of Problem \ref{p:b_and_r} can be evaluated by:
\begin{prob} \label{p:b_and_r_evaluate}
	\begin{subequations}
		\begin{align}
		\mathcal{G} \left( \lambda, \varepsilon \right) = \min_{z, \{r_n, b_n|n \in \mathcal{N}\}} & \mathcal{L} \left( z, \{r_n, b_n|n \in \mathcal{N}\}, \lambda, \varepsilon \right) \\
		\text{s.t.} \qquad & (\ref{e:first_con_r}), (\ref{e:first_con_b})
		\end{align}
	\end{subequations}
\end{prob}
And the dual problem of Problem \ref{p:b_and_r} is given as:
\begin{prob} \label{p:b_and_r_dual}
	\begin{subequations}
		\begin{align}
		\max_{\lambda, \varepsilon} \quad & \mathcal{G} \left( \lambda, \varepsilon \right) \\
		\text{s.t.} \quad & \lambda \geq 0, \varepsilon \geq 0
		\end{align}
	\end{subequations}
\end{prob}

Since Problem \ref{p:b_and_r} is convex and satisfies Slater's condition, strong duality holds between the primary problem and the dual problem, i.e. Problem \ref{p:b_and_r} and Problem \ref{p:b_and_r_dual}. 
Thus solving Problem \ref{p:b_and_r_dual} is equivalent with solving Problem \ref{p:b_and_r}.
To this end, Problem \ref{p:b_and_r} can be solved by first evaluating the Lagrangian with fixed $\lambda$ and $\varepsilon$, then searching for the optimal $\lambda$ and $\varepsilon$, which will be discussed in the following two parts respectively.

\subsubsection{Evaluation of the dual function}
Regarding the evaluation of the Lagrangian, 
let ${[\cdot]}^*$ denote the optimal solution of variable $[\cdot]$,
the following lemma can be expected.

\begin{lem} \label{lem:optimal_x}
	Define function $f(x) = (t_1 x -t_2) e^{\frac{t_1 x}{t_2}} +t_2$. 
	The optimal solution of $\{r_n\}$, $\{b_n\}$ and $z$ can be given as follows:
	\begin{align} 
	&{r_n}^* = \max\left\{ 0, 1 - \sqrt{
		\frac{\frac{t_1 c_n}{h_n} e^{\frac{t1 f^{-1} \left(\lambda h_n \right)}{t_2}} + \varepsilon}{3 \kappa_n t_1 X^3 c_n^3}
	}\right\}, \forall n \in \mathcal{N} \label{e:optimal_r} \\
	&{b_n}^* = \left \{ \begin{array}{ll}
	\frac{c_n {r_n}^*}{f^{-1} \left(\lambda h_n \right)},  {r_n}^* = 1 - \sqrt{
		\frac{\frac{t_1 c_n}{h_n} e^{\frac{t1 f^{-1} \left(\lambda h_n \right)}{t_2}} + \varepsilon}{3 \kappa_n t_1 X^3 c_n^3}} > 0 \\
	0,  {r_n}^* = 0 \end{array} \right. , \forall n \in \mathcal{N}  \label{e:optimal_b}\\
	&z^* = \frac{t_3 B}{t_1} \ln \frac{\varepsilon g}{t_1} \label{e:optimal_z}
	\end{align}
\end{lem}
\begin{IEEEproof}
	Before going into details of the proof, it should be affirmed that:
	For $\forall n \in \mathcal{N}$, ${r_n}^*$ and ${b_n}^*$ should be 0 or not 0 simultaneously, i.e. ${r_n}^* = 0$ is sufficient and necessary condition of ${b_n}^* = 0$.
	The reason for this property can be interpreted by resorting to the physical meaning of ${r_n}$ and ${b_n}$.
	First we explain the sufficiency of the condition, i.e. `if ${r_n}^*=0$, then ${b_n}^*=0$'.
	If the offloading ratio of user $n$ is $r_n = 0$, then no transmission happens between user $n$ and the cooperative node.
	Therefore the bandwidth should be allocated to other users to save energy for their transmission, thus $b_n$ should be 0.
	The interpretation of the necessity of the condition, i.e. `if ${b_n}^*=0$, then ${r_n}^*=0$', is similar and omitted here.
	
	With the above result, we investigate the partial derivative of the Lagrangian:	
	\begin{align}
	\frac{\partial \mathcal{L}}{\partial b_n} \bigg|_{b_n ={b_n}^*} =
	& -\frac{t_1 c_n r_n}{b_n h_n} e^{\frac{t1 c_n r_n}{t_2 b_n}} + \frac{t_2}{h_n} \left( e^{\frac{t_1 c_n r_n}{t_2 b_n}} -1 \right) +\lambda \bigg|_{b_n ={b_n}^*} \notag \\
	& \left \{ \begin{array}{ll}
	=0,  {b_n}^* > 0 \\
	\geq 0,  {b_n}^* = 0
	\end{array} \right. \label{e:partial_derivative_b} \\
	\frac{\partial \mathcal{L}}{\partial r_n} \bigg|_{r_n ={r_n}^*} =
	& \frac{t_1 c_n}{h_n} e^{\frac{t1 c_n r_n}{t_2 b_n}} - 3 \kappa_n t_1 X^3 c_n^3 \left( 1-r_n \right)^2 + \varepsilon \bigg|_{r_n ={r_n}^*} \notag \\
	& \left \{ \begin{array}{ll}
	\geq 0,  {r_n}^* = 0 \\
	=0,  0 < {r_n}^* < 1 \\
	\leq 0,  {r_n}^* = 1
	\end{array} \right. \label{e:partial_derivative_r} \\
	\frac{\partial \mathcal{L}}{\partial z} \bigg|_{z =z^*} =
	& \frac{t_1}{g} e^{\frac{t_1 z}{t_3 B}} - \varepsilon \bigg|_{z =z^*} =0\label{e:partial_derivative_z}
	\end{align}
	
	First we look into (\ref{e:partial_derivative_b}). Let $\alpha_n = \frac{c_n r_n}{b_n}$. When ${b_n}^* > 0$, it is certain that ${r_n}^* > 0$, equation (\ref{e:partial_derivative_b}) can be transformed into
	\begin{equation}
	\lambda = \frac{t_1 c_n r_n}{b_n h_n} e^{\frac{t1 c_n r_n}{t_2 b_n}} - \frac{t_2}{h_n} \left( e^{\frac{t_1 c_n r_n}{t_2 b_n}} -1 \right)
	\end{equation}
	which indicates that
	\begin{equation}
	\lambda h_n = f(\alpha_n)
	\end{equation}
	Retrospecting that $t_1$ should be larger than $t_2$, by checking the first-order partial derivative, it is straightforward that $f(x)$ is a monotonic increasing function, which is thus reversible.
	To this end, define $\mathcal{N}_0 = \{ n|r_n = 0 \}$ and $\mathcal{N}_1 = \{ n|r_n > 0 \}$, it can be seen that the optimal solution of ${r_n}^*$ and ${b_n}^*$ in Problem \ref{p:b_and_r_evaluate} are related such that
	\begin{equation} \label{e:optimal_b0}
		{b_n}^* = 0, n \in \mathcal{N}_0
	\end{equation}
	and
	\begin{equation} \label{e:optimal_alpha}
		\frac{c_n r_n^*}{b_n^*}=f^{-1} \left(\lambda h_n \right), n \in \mathcal{N}_1
	\end{equation}
	Besides, it directly follows that
	\begin{align}
		\mathcal{N}_0 \cup \mathcal{N}_1 &= \mathcal{N} \notag \\
		\mathcal{N}_0 \cap \mathcal{N}_1 &= \emptyset
	\end{align}
	
	Next, we investigate (\ref{e:partial_derivative_r}).
	Above all, since the left-hand side of (\ref{e:partial_derivative_r}) is larger than 0 at $r_n = 1$, 
	it is certain that the optimal solution of $r_n$ satisfies ${r_n}^* < 1$ for $\forall n \in \mathcal{N}$. The optimal solution can be expressed within set $\mathcal{N}_1$ and $\mathcal{N}_0$ respectively:
	\begin{itemize}
		\item 
		For $n \in \mathcal{N}_1$, the equation of (\ref{e:partial_derivative_r}) holds. 
		Substitute (\ref{e:optimal_alpha}) into the first term of (\ref{e:partial_derivative_r}), it is expected that 
		\begin{equation} \label{e:optimal_r_n1}
			{r_n}^* =  1 - \sqrt{
				\frac{\frac{t_1 c_n}{h_n} e^{\frac{t1 f^{-1} \left(\lambda h_n \right)}{t_2}} + \varepsilon}{3 \kappa_n t_1 X^3 c_n^3}}, \forall n \in \mathcal{N}_1
		\end{equation}
		The optimal solution of $b_n$ can be expressed by substituting (\ref{e:optimal_r_n1}) into (\ref{e:optimal_alpha}) that 
		\begin{equation} \label{e:optimal_b_n1}
			{b_n}^* = \frac{c_n {r_n}^*}{f^{-1} \left(\lambda h_n \right)}, \forall n \in \mathcal{N}_1
		\end{equation}
		\item 
		For $n \in \mathcal{N}_0$, it is straightforward that
		\begin{equation} \label{e:optimal_rb_n0}
			{r_n}^* = {b_n}^* = 0, \forall n \in \mathcal{N}_0
		\end{equation}
	\end{itemize}
	Combining the conclusions in (\ref{e:optimal_r_n1}), (\ref{e:optimal_b_n1}) and (\ref{e:optimal_rb_n0}), the expression in (\ref{e:optimal_r}) and (\ref{e:optimal_b}) can be given out.
	
	Finally, leveraging (\ref{e:partial_derivative_z}), (\ref{e:optimal_z}) is derived.

	This completes the proof.
\end{IEEEproof}

\subsubsection{Solution of the dual variables}
\sloppy
In the previous subsection, the optimal solution of $\left \{ z, \{r_n, b_n|n \in \mathcal{N}\} \right \}$ is given out by two lagrangian multipliers $\lambda$ and $\varepsilon$, i.e. $\left \{ z^*, \{{r_n}^*, {b_n}^*|n \in \mathcal{N}\} \right \}$ can be recognized as functions of $\lambda$ and $\varepsilon$.
To this end, these two variables generally should be searched with subgradient-based method in convex optimization \cite{convex_book}, which, however, can be very slow sometimes. Besides, the choice on step size and stopping criterion of subgradient method can be very tricky \cite{numerical_book}.
By utilizing the optimal analytical structure of the primal variables, we will find the solution of the two multipliers with a bilevel bisection search, as introduced with the following two lemmas.

\begin{lem} \label{lem:search_epsilon}
	With fixed $\lambda$, the solution of $\varepsilon$ can be obtained by a single-variable bisection search.
\end{lem}
\begin{IEEEproof}
	As introduced in Section \ref{s:first_problem_bandr}, for the optimal solution of Problem \ref{p:b_and_r}, equation $\sum_{n} c_n r_n^* = z^*$ should satisfy.
	With given $\lambda$, on the left-hand side of this equation, seen from (\ref{e:optimal_r}), variable $r_n$ is a decreasing function of $\varepsilon$ before reaching 0.
	On the right-hand side of this equation, variable $z$ is an increasing function of $\varepsilon$, as shown in (\ref{e:optimal_z}).
	Thus there is only one solution to $\varepsilon$ that can achieve the optimal solution of $r_n$ satisfying the equation of $\sum_{n} c_n r_n^* = z^*$, which can be find out by a bisection search. 
	
	This completes the proof.
\end{IEEEproof}

With Lemma \ref{lem:search_epsilon}, for arbitrary given $\lambda \geq 0$, a solution of $\varepsilon$ can be given out, which can be recognized as a lower level search.
After executing this lower level search, ${r_n}^*$ and ${b_n}^*$ can be recognized as functions of $\lambda$.
In the upper level, we search for the optimal value of $\lambda$ by resorting to the expression of ${b_n}^*$ in (\ref{e:optimal_b}).
To be specific, the following lemma can be established.
\begin{lem} \label{lem:search_lambda}
	The solution of $\lambda$ can be obtained by a single-variable bisection search.
\end{lem}
\begin{IEEEproof}
	Before going into details of the proof, the following result should be established:
	For the optimal solution of Problem \ref{p:b_and_r}, there is $\sum_{n} {b_n}^* = B$.
	This result is quite straightforward by referring to the physical meaning.
	Since the energy consumption of wireless transmission is monotonic decreasing with transmit bandwidth, the overall transmit bandwidth should be as large as possible in order to save energy.
	Thus the users should make full use of the system bandwidth, therefore $\sum_{n} {b_n}^* = B$ should satisfy.

	Next we explore the monotonic property of ${r_n}^*$ with respect to $\lambda$ when $0<{r_n}^*<1$.
	In this case, sum up equation (\ref{e:partial_derivative_r}) and (\ref{e:partial_derivative_z}), there is
	\begin{equation} \label{e:opt_lambda_original}
	\frac{t_1 c_n}{h_n} e^{\frac{t1 c_n {r_n}^*}{t_2 {b_n}^*}}- 3 \kappa_n t_1 X^3 c_n^3 \left( 1-{r_n}^* \right)^2 + \frac{t_1}{g} e^{\frac{t_1 z^*}{t_3 B}} = 0
	\end{equation}
	Substituting (\ref{e:optimal_alpha}) and $z^* = \sum_{n} {r_n}^*$ into the first and third term, respectively. (\ref{e:opt_lambda_original}) is transformed into
	\begin{equation} \label{e:opt_lambda_final}
	\frac{t_1 c_n}{h_n} e^{\frac{t1 f^{-1} \left(\lambda h_n \right)}{t_2}}- 3 \kappa_n t_1 X^3 c_n^3 \left( 1-{r_n}^* \right)^2 + \frac{t_1}{g} e^{\frac{t_1 \sum_{n} {r_n}^*}{t_3 B}} = 0
	\end{equation}
	In (\ref{e:opt_lambda_final}), the first term is monotonic increasing function of $\lambda$ and the last two terms are increasing functions of ${r_n}^*$. 
	Note that ${r_n}^*(\lambda)$ should be the solution of (\ref{e:opt_lambda_final}).
	When increasing $\lambda$, the first term is increased and ${r_n}^*(\lambda)$ should be smaller in order to meet the equation.
	Thus ${r_n}^*(\lambda)$ is a decreasing function of $\lambda$.
	Retrospecting in Lemma \ref{lem:optimal_x} that
	\begin{align}
	b_n^*(\lambda) = \frac{c_n r_n^*(\lambda)}{f^{-1} \left(\lambda h_n \right)} 
	\end{align}
	Since ${r_n}^*(\lambda)$ and $f^{-1} (\lambda h_n)$ are decreasing and increasing function of $\lambda$, respectively, ${b_n}^*(\lambda)$ is decreasing with $\lambda$.
	Therefore, there is only one solution of $\lambda$ that can achieve $\sum_{n} {b_n}^*(\lambda) = B$, which can be obtained by a bisection search due to the monotonic property \footnote{As for $n \in \mathcal{N}$ that ${r_n}^* = {b_n}^* = 0$, they does not interfere with the term $\sum_{n} {b_n}^*$, and thus not discussed specifically.}.
	
	This completes the proof.
\end{IEEEproof}

\subsection{Optimal time allocation} \label{s:first_problem_t}
Finally, we turn to find the optimal time allocation $t_1, t_2, t_3$ with fixed $\{r_n, b_n|n \in \mathcal{N}\}$, whose relative optimization problem is
\begin{prob} \label{p:optimize_t}
	\begin{subequations}
		\begin{align}
		\min_{\substack{t_1, t_2, t_3}}
		& {\sum_{n} \frac{t_2 b_n}{h_n} \left( e^{\frac{t_1 c_n r_n}{t_2 b_n}} -1 \right) + \frac{t_3 B}{g} \left( e^{\frac{t_1 \sum_{n} c_n r_n}{t_3 B}} -1 \right) +\sum_{n} \kappa_n t_1 X^3 {c_n}^3 {(1-r_n)}^3} \notag \\
		& + \sum_{n} E^o_n - \theta {\left(t_1 +t_2 +t_3\right)} \notag \\
		\text{s.t.} \quad
		& (\ref{e:first_con_time_lower_upper}),(\ref{e:first_con_time}).
		\end{align}
	\end{subequations}
\end{prob}

In the objective function of Problem \ref{p:optimize_t}, since the perspective function of expression function is still convex \cite{convex_book}, the first and second term are joint convex functions, whereas the last two terms are linear with respect to $t_1, t_2, t_3$. 
Besides, constraint (\ref{e:first_con_time}) is linear.
Therefore, Problem \ref{p:optimize_t} is a convex problem with its solution able to be obtained directly by interior-point method.

\section{Solution for the case when the BS has finite computation capacity} \label{s:optimal_solution_second}

In this section, we will find the local optimal solution Problem \ref{p:second} and discuss the convergence of the proposed algorithm.

\subsection{Problem solution}
Since the objective function of Problem \ref{p:second} is in a fractional structure as well, following similar discussion as in Section \ref{s:first_problem_transformation}, Problem \ref{p:second} can be solved by iteratively executing the following two steps given an initial value of $\vartheta$.

1) Solve Problem \ref{p:second_minus} with fixed $\vartheta$:
\begin{prob} \label{p:second_minus}
	\begin{subequations}
		\begin{align}
			\min_{\substack{t_1, t_2, t_3, t_4, \\ \{r_n, b_n|n \in \mathcal{N}\}}}
			& {\sum_{n} \frac{t_2 b_n}{h_n} \left( e^{\frac{t_1 c_n r_n}{t_2 b_n}} -1 \right) + \frac{t_3 B}{g} \left( e^{\frac{t_1 \sum_{n} c_n r_n}{t_3 B}} -1 \right) +\sum_{n} \kappa_n t_1 X^3 {c_n}^3 {(1-r_n)}^3} \notag \\
			& + \sum_{n} E^o_n - \vartheta {\left(t_1 +t_2 +t_3 +t_4\right)} \notag \\
			\text{s.t.} \quad
			& (\ref{e:second_con_time_lower_upper})-(\ref{e:second_con_b}).
		\end{align}
	\end{subequations}
\end{prob}

2) Let $\vartheta = \frac{{\sum_{n} \frac{t_2 b_n}{h_n} \left( e^{\frac{t_1 c_n r_n}{t_2 b_n}} -1 \right) + \frac{t_3 B}{g} \left( e^{\frac{t_1 \sum_{n} c_n r_n}{t_3 B}} -1 \right) +\sum_{n} \kappa_n t_1 X^3 {c_n}^3 {(1-r_n)}^3} + \sum_{n} E^o_n}{t_1 +t_2 +t_3 +t_4}$

Problem \ref{p:second_minus} is still not tractable since the variable $t_1$ is coupled with $\{r_n\}$ and $t_2$ is coupled with $\{b_n\}$.
To this end, we define $p_n = t_2 b_n$ and $q_n = t_1 r_n$ for $n \in \mathcal{N}$ and substitute $b_n$ and $r_n$ with $p_n/t_2$ and $q_n/t_1$, respectively. 
Following the physical definition, $t_1, t_2$ cannot take the value 0, thus the special case when $b_n = 0$ or $r_n = 0$ can be directly replaced by $p_n = 0$ or $q_n = 0$.
Problem \ref{p:second_minus} is equivalently transformed into

\begin{prob} \label{p:second_pq}
	\begin{subequations}
		\begin{align}
			\min_{\substack{t_1, t_2, t_3, t_4, \\ \{q_n, p_n|n \in \mathcal{N}\}}}
			& \gamma \left(t_1, t_2, t_3, t_4, \{q_n, p_n|n \in \mathcal{N}\}\right) \notag \\
			= \quad & \sum_{n} \frac{p_n}{h_n} \left( e^{\frac{c_n q_n}{p_n}} -1 \right) + \frac{t_3 B}{g} \left( e^{\frac{\sum_{n} c_n q_n}{t_3 B}} -1 \right) +\sum_{n} \kappa_n X^3 {c_n}^3 \left(t_1 + \frac{3 q_n^2}{t_1} - \frac{q_n^3}{t_1^2} -3 q_n\right) \notag \\
			& + \sum_{n} E^o_n - \vartheta {\left(t_1 +t_2 +t_3 +t_4\right)} \notag \\
			\text{s.t.} \quad
			& (\ref{e:second_con_time_lower_upper}),(\ref{e:second_con_time}) \\
			& \sum_{n = 1}^N X c_n q_n \leq f_B t_4, \label{e:second_minus_con_comp} \\
			& \sum_{n = 1}^N p_n \leq B t_2,  \label{e:second_minus_con_band} \\
			& 0 \leq q_n \leq t_1, n \in \mathcal{N}, \label{e:second_minus_con_r}\\
			& p_n \geq 0, n \in \mathcal{N}. \label{e:second_minus_con_b}.
		\end{align}
	\end{subequations}
\end{prob}
In Problem \ref{p:second_pq}, (\ref{e:second_minus_con_comp})-(\ref{e:second_minus_con_b}) is derived from (\ref{e:second_con_comp})-(\ref{e:second_con_b}).
The constraints of Problem \ref{p:second_pq} are linear, but the objective function is nonconvex whose global optimal solution is still hard to find.
However, the objective function of Problem \ref{p:second_pq} can be recognized as difference of two jointly convex functions of ${t_1, t_2, t_3, t_4, \{q_n, p_n|n \in \mathcal{N}\}}$, i.e. $\gamma = \psi - \eta$, where
$\psi = \sum_{n} \frac{p_n}{h_n} \left( e^{\frac{c_n q_n}{p_n}} -1 \right) + \frac{t_3 B}{g} \left( e^{\frac{\sum_{n} c_n q_n}{t_3 B}} -1 \right) +\sum_{n} \kappa_n X^3 {c_n}^3 \left(t_1 + \frac{3 q_n^2}{t_1}\right) + \sum_{n} E^o_n$ and $\eta = \sum_{n} \kappa_n X^3 {c_n}^3 \left(\frac{q_n^3}{t_1^2} +3 q_n\right) + \vartheta {\left(t_1 +t_2 +t_3 +t_4\right)}$.
Therefore, Problem \ref{p:second_pq} is a standard difference of convex functions programming (DCP) and can be solved by difference of convex algorithms (DCA).

The general idea of DCA is to express the objective function as difference of two convex functions and update the two parts in separate steps \cite{dca_paper}.
Suppose the objective function can be expressed as $\psi([\cdot]) -\eta([\cdot])$, denote the iteration index as $k$ and initialize $k = 0$, $x^{(0)} \in \text{dom } \psi$.
In each iteration, repeat the following two steps within the domain of the optimization problem.
\begin{enumerate}
	\item With fixed $x^{(k)} \in \text{dom } \psi$, 
	find $y^{(k)} \in \partial \eta(x^{(k)})$.
	\item With fixed $y^{(k)} \in \text{dom } \eta$, 
	solve $x^{(k+1)} = \arg \min_x \left\{ \psi(x) - \left[ \eta(x^{(k)}) + y^{(k)} (x - x^{(k)}) \right] \right\}$.
\end{enumerate}
In the above step 1), $\partial \eta(x^{(k)})$ is the subgradient of $\eta([\cdot])$ 
at $x^{(k)}$.
In step 2), the objective function to be minimized is the difference of $\psi([\cdot])$ and the first-order approximate of $\eta([\cdot])$ at $x^{(k)}$ with subgradient $y^{(k)}$.

\sloppy
Looking into Problem \ref{p:second_pq}, thanks to the fact that function $\eta(t_1, t_2, t_3, t_4, \{q_n, p_n|n \in \mathcal{N}\})$ is continuous and convex function, the subgradient of $\eta$ can be directly replaced by its gradient, which no longer requires further optimization.
Thus the procedure for finding the optimal solution of Problem \ref{p:second_pq} is reduced into simply repeating step 2) in the above.
At the $k$-th iteration, we replace function $\eta$ with its first-order approximation at $(t_1^{(k)}, t_2^{(k)}, t_3^{(k)}, t_4^{(k)}, \{q_n^{(k)}, p_n^{(k)}|n \in \mathcal{N}\})$ and find the optimal solution of this approximation problem, which is set to be $(t_1^{(k+1)}, t_2^{(k+1)}, t_3^{(k+1)}, t_4^{(k+1)}, \{q_n^{(k+1)}, p_n^{(k+1)}|n \in \mathcal{N}\})$.
By iteratively repeating this step from a feasible point $(t_1^{(0)}, t_2^{(0)}, t_3^{(0)}, t_4^{(0)}, \{q_n^{(0)}, p_n^{(0)}|n \in \mathcal{N}\})$ of Problem \ref{p:second_pq}, convergent solution of Problem \ref{p:second_pq} can be found.

Define the first-order approximate of $\eta$ at $(t_1^{(k)}, t_2^{(k)}, t_3^{(k)}, t_4^{(k)}, \{q_n^{(k)}, p_n^{(k)}|n \in \mathcal{N}\})$ is $\overline{\eta}^{(k)}$, then
{
	\small
	\begin{align}
	& \overline{\eta}^{(k)}(t_1, t_2, t_3, t_4, \{q_n, p_n|n \in \mathcal{N}\}) \notag \\
	= \quad & \eta(t_1^{(k)}, t_2^{(k)}, t_3^{(k)}, t_4^{(k)}, \{q_n^{(k)}, p_n^{(k)}|n \in \mathcal{N}\}) + \sum_{n} \frac{\partial \eta}{\partial q_n} \bigg|_{q_n = q_n^{(k)}} \left( q_n - q_n^{(k)} \right)
	+ \sum_{i=1}^4 \frac{\partial \eta}{\partial t_i} \bigg|_{t_i = t_i^{(k)}} \left( t_i - t_i^{(k)} \right) \notag \\
	= \quad & \sum_{n} \kappa_n X^3 {c_n}^3 \left(\frac{{q_n^{(k)}}^3}{{t_1^{(k)}}^2} +3 q_n^{(k)} \right) + \vartheta {\left(t_1^{(k)} +t_2^{(k)} +t_3^{(k)} +t_4^{(k)} \right)} + \sum_{n} k_n X^3 {c_n}^3 \left( \frac{3 {q_n^{(k)}}^2}{{t_1^{(k)}}^2}+3 \right) \left( q_n - q_n^{(k)} \right) \notag \\ 
	\quad & + \left( \vartheta - \frac{2 X^3}{{t_1^{(k)}}^3} \sum_{n} \kappa_n {c_n}^3 {q_n^{(k)}}^3 \right) \left(t_1 - t_1^{(k)}\right) + \vartheta \left(t_2 - t_2^{(k)}\right) + \vartheta \left(t_3 - t_3^{(k)}\right) + \vartheta \left(t_4 - t_4^{(k)}\right) \notag \\
	= \quad & \sum_{n} \kappa_n X^3 {c_n}^3 \left(\frac{{q_n^{(k)}}^3}{{t_1^{(k)}}^2} +3 q_n^{(k)} \right) + \vartheta {\left(t_1 +t_2 +t_3 +t_4 \right)} \notag \\
	\quad & + \sum_{n} k_n X^3 {c_n}^3 \left( \frac{3 {q_n^{(k)}}^2}{{t_1^{(k)}}^2}+3 \right) \left( q_n - q_n^{(k)} \right) - \frac{2 X^3}{{t_1^{(k)}}^3} \left(t_1 - t_1^{(k)}\right) \sum_{n} \kappa_n {c_n}^3 {q_n^{(k)}}^3
	\end{align}
}
The optimization problem to be solved in the $k$-th iteration is given as
\begin{prob} \label{p:second_pq_iterative}
	\begin{subequations}
		\begin{align}
			\min_{\substack{t_1, t_2, t_3, t_4, \\ \{q_n, p_n|n \in \mathcal{N}\}}}
			& \overline{\gamma}^{(k)}(t_1, t_2, t_3, t_4, \{q_n, p_n|n \in \mathcal{N}\}) \notag \\
			& = \psi(t_1, t_2, t_3, t_4, \{q_n, p_n|n \in \mathcal{N}\}) - \overline{\eta}^{(k)}(t_1, t_2, t_3, t_4, \{q_n, p_n|n \in \mathcal{N}\}) \notag \\
			\text{s.t.} \quad
			& (\ref{e:second_con_time_lower_upper}), (\ref{e:second_con_time}), (\ref{e:second_minus_con_comp}), (\ref{e:second_minus_con_band}), (\ref{e:second_minus_con_r}), (\ref{e:second_minus_con_b}).
		\end{align}
	\end{subequations}
\end{prob}
Since $\psi$ and $\eta^{(k)}$ are convex and linear functions with respect to ${t_1, t_2, t_3, t_4, \{q_n, p_n|n \in \mathcal{N}\}}$, respectively, and the constraints of Problem \ref{p:second_pq_iterative} are linear, Problem \ref{p:second_pq_iterative} is a convex problem which can be effectively solved by interior-point method.

\subsection{Convergence analysis} \label{s:case_2_convergence_analysis}
In the previous subsection, the DCA algorithm for solving Problem \ref{p:second_pq} is presented.
The convergence of this algorithm is discussed in this subsection.
For $k$-th iteration with fixed point $(t_1^{(k)}, t_2^{(k)}, t_3^{(k)}, t_4^{(k)}, \{q_n^{(k)}, p_n^{(k)}|n \in \mathcal{N}\})$, the derived optimal solution by solving Problem \ref{p:second_pq_iterative} is set as $(t_1^{(k+1)}, t_2^{(k+1)}, t_3^{(k+1)}, t_4^{(k+1)}, \{q_n^{(k+1)}, p_n^{(k+1)}|n \in \mathcal{N}\})$.
In this subsection, to simplify the expression, we denote the point $(t_1, t_2, t_3, t_4, \{q_n, p_n|n \in \mathcal{N}\})$ as $\boldsymbol{y}$, and $(t_1^{(k)}, t_2^{(k)}, t_3^{(k)}, t_4^{(k)}, \{q_n^{(k)}, p_n^{(k)}|n \in \mathcal{N}\})$ as $\boldsymbol{y}^{(k)}$.
Looking into Problem \ref{p:second_pq} and Problem \ref{p:second_pq_iterative}, they have different objective function but identical feasible region,
thus $\boldsymbol{y}^{(k+1)}$ is also in the feasible region of Problem \ref{p:second_pq}.
Denote the objective function value of Problem \ref{p:second_pq} at $\boldsymbol{y}^{(k+1)}$ as $\Gamma^{(k+1)}$, and the optimal value of Problem \ref{p:second_pq_iterative} at $k$-th iteration, which is related to optimal solution $\boldsymbol{y}^{(k+1)}$, as $\overline{\Gamma}^{(k+1)}$,
the following lemma can be established.

\begin{lem} \label{lem:dca_convergence}
	From arbitrary $\boldsymbol{y}^{(0)}$ in the feasible region of Problem \ref{p:second_pq}, the proposed DCA algorithm generates a sequence of $\overline{\Gamma}^{(k)}$, which is decreasing and will converge to a stationary point.
\end{lem}
\begin{IEEEproof}
	Since $\eta$ is a jointly convex function, there is
	\begin{equation}
		\eta(\boldsymbol{y}) \geq \overline{\eta}^{(k)}(\boldsymbol{y})
	\end{equation}
	So it is obvious that
	\begin{equation} \label{e:gamma_inequality}
		\gamma(\boldsymbol{y}) \leq \overline{\gamma}^{(k)}(\boldsymbol{y})
	\end{equation}
	Recollecting that the optimal point of Problem \ref{p:second_pq_iterative} at $\boldsymbol{y}^{(k)}$ is $\boldsymbol{y}^{(k+1)}$.
	It directly follows that
	\begin{align} \label{e:dca_convergence_better}
		& ~\Gamma^{(k+1)} = \gamma(\boldsymbol{y}^{(k+1)}) \notag \\
		\leq & ~\overline{\Gamma}^{(k+1)} = \overline{\gamma}^{(k)} (\boldsymbol{y}^{(k+1)}) \notag \\
		\leq & ~\overline{\gamma}^{(k)} (\boldsymbol{y}^{(k)}) \notag \\
		= & ~\Gamma^{k} = \gamma (\boldsymbol{y}^{(k)})
	\end{align}
	In (\ref{e:dca_convergence_better}), the inequality on the second line is expected since (\ref{e:gamma_inequality}) holds in the whole feasible region, especially at $\boldsymbol{y}^{(k+1)}$. The inequality on the third line is established because $\boldsymbol{y}^{(k+1)}$ is the optimal solution of Problem \ref{p:second_pq_iterative} at the $k$-th iteration. The equation on the fourth line holds by definition of function $\overline{\gamma}^{(k)}$.
		
	The inequality in (\ref{e:dca_convergence_better}) indicates that $\boldsymbol{y}^{(k+1)}$ yields lower cost in terms of cost function of Problem \ref{p:second_pq} than $\boldsymbol{y}^{(k)}$. From Cauchy's theorem, in the sequence $\big \{ \boldsymbol{y}^{(k)} \big \}$, there is a convergent subsequence $\big \{\boldsymbol{y}^{(k_v)} \big \}$ with limit point $\boldsymbol{y}^{*}$, such that
	$\lim_{v\to \infty} \left( \gamma ( \boldsymbol{y}^{(k_v)} ) - \gamma \left(\boldsymbol{y}^{*} \right) \right) =0.$
	For certain $k$, there must exists a $v$ satisfiying $k_v \leq k \leq k_{v+1}$, hence there is
	$\gamma \left( \boldsymbol{y}^{(k_v)} \right)
	\geq \gamma \left( \boldsymbol{y}^{(k)} \right)
	\geq \gamma \left( \boldsymbol{y}^{(k_{v+1})} \right).$
	When $k$ goes to infinity, we have
	\begin{equation}
		\begin{split}
			0=&\lim_{v\to \infty} \left(\gamma ( \boldsymbol{y}^{(k_v)} ) - \gamma ( \boldsymbol{y}^{*} ) \right) \\
			\geq & \lim_{i\to \infty} \left(\gamma ( \boldsymbol{y}^{(k)} ) - \gamma ( \boldsymbol{y}^{*} ) \right) \\
			\geq & \lim_{v\to \infty} \left(\gamma ( \boldsymbol{y}^{(k_{v+1})} ) - \gamma ( \boldsymbol{y}^{*} ) \right) =0.
		\end{split}
	\end{equation}
	Therefore, the sequence $\left( \boldsymbol{y}^{(k)} \right)$ is also convergent, with limit point being $\lim_{k\to \infty} \left( \boldsymbol{y}^{(k)} \right)= \boldsymbol{y}^{*} $. 
	Note that the feasible region of Problem \ref{p:second_pq} is closed, it directly follows that $\boldsymbol{y}^{*}$ falls in the feasible region and would be a stationary point.
\end{IEEEproof}

\section{Numerical Results} \label{s:numerical_results}
In this section, numerical results of our proposed algorithms for Problem \ref{p:first} and Problem \ref{p:second} are presented.
The default system parameters are set as follows.
There are $N=5$ users in the system, whose distance from the cooperative node follows a uniform distribution in the interval [5m,50m].
The distance between the cooperative node and the BS is 30m.
The overall bandwidth in the system is $B=1$ MHz. 
The channels experiences distance attenuation and Rayleigh fading,
hence the uniformed channel gain (i.e. uniformed signal-to-noise ratio) is the product of these two components divided by the power spectrum density of noise.
In terms of the distance attenuation, the pathloss is denoted as $g_0\left(d/d_0\right)^{-3}$, 
where $g_0=-60$ dB corresponds to the path loss at the reference distance of $d_0 = 10$m, $d$ denotes the distance from the transmitter to the receiver.
The random gain under Rayleigh fading obeys exponential distribution with mean being 1.
Considering the noise both in the environment and the receiver, the power spectrum density of noise is set as $\sigma^2=-110$ dbm by default \footnote{In the following, the power of noise will be referred to with unit of mW unless special statement}.
Similar to \cite{ref_97}, the coefficient for local computing $\kappa=10^{-24}$. 
Suppose the users generate computation tasks with images, which are generally large in data size and heavy in computation burden,
the data accumulated per second $c_n = 1.5 \times 10^6$ nat and the required computation capacity to compute unit nat of task $X = 100$ \cite{Self_dependency}.
The computation capacity of the BS is $f_B=5$ GHz.
The numerical simulation was executed using Matlab 2019b on a Windows 10 platform with an Intel Core i7-6700 processor and 16GB RAM.

\subsection{Complexity analysis of the proposed method for Case I}

\begin{figure}
	\begin{center}
		\includegraphics[angle=0,width=0.42 \textwidth]{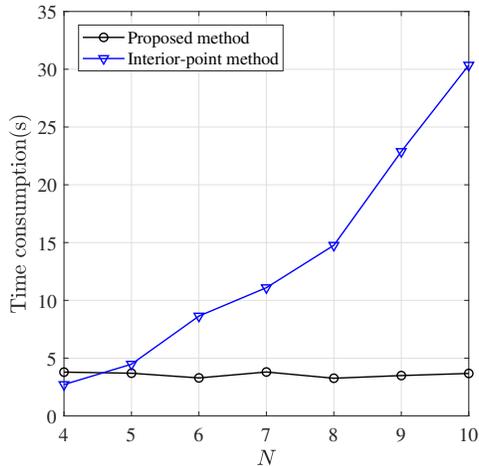}
	\end{center}
	\caption{Comparison of the time consumption for solving Problem \ref{p:first_minus} in Case I.}
	\label{f:case_1_complexity}
\end{figure}

In this subsection, the computational complexity of our proposed method for solving Problem \ref{p:first_minus} is analyzed.
Problem \ref{p:first_minus} is a convex problem for $\{r_n, b_n|n \in \mathcal{N}\}$ when $t_1,t_2,t_3$ are fixed, thus can be solved by existing algorithms optimally.
As a comparison of the proposed method, the computational complexity for solving Problem \ref{p:first_minus} via interior-point method, 
which is one of the most popular numerical methods for solving a convex problem optimally, is evaluated.

In Fig. \ref{f:case_1_complexity}, the accumulated time consumption for solving Problem \ref{p:first_minus} with respect to $\{r_n, b_n|n \in \mathcal{N}\}$ over 100 sets of randomly generated channel gains is depicted with user number $N$.
Specifically, for different value of $N$, distance between each user and the cooperative node, along with the Rayleigh fading coefficient within the link of each user, are selected from the default distribution. Other parameters are set as described above in the default settings.
At $N=4$, the time consumption for solving Problem \ref{p:first_minus} by interior-point method is lower than the time consumption of the proposed method, who mostly lag behind due to the delay in the transformation between the primary variables and the dual variables.
On the other hand, since increasing $N$ implicates that more optimization variables and more constraints need to be considered, 
the time consumption for solving Problem \ref{p:first_minus} by interior-point method shows a significant increasing tendency with $N$ in Fig. \ref{f:case_1_complexity}.
But the time consumption of the proposed method hardly change with different $n$, 
which confirms the superiority of our proposed compared to conventional methods,
benefiting from the fact that the proposed method always reach the optimal solution by a two-layer bisection search.

\subsection{Convergence of the proposed DCA algorithm for Case II}

\begin{figure}
	\begin{center}
		\includegraphics[angle=0,width=0.42 \textwidth]{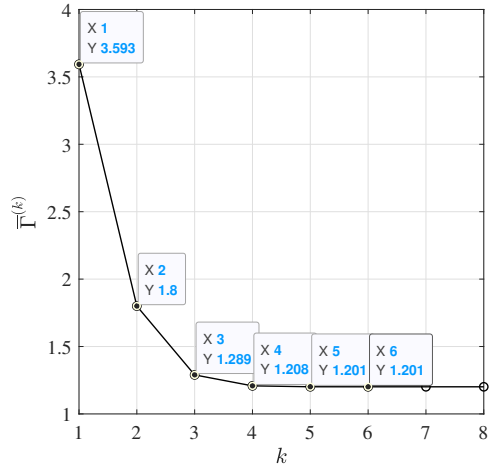}
	\end{center}
	\caption{Convergence of DCA algorithm in Case II.}
	\label{f:case_2_convergence}
\end{figure}

In Section \ref{s:case_2_convergence_analysis}, the convergence of the proposed DCA algorithm is proved.
In order to verify this result, the convergence speed of DCA algorithm with the default parameters is illustrated in Fig. \ref{f:case_2_convergence}.
Checking the function value within each iteration, as the iteration index $k$ increases, the value of $\overline{\Gamma}^{(k)}$ initially experiences a sharp drop, 
which later levels off but continues to decrease gradually until it reaches a tolerance of no more than 0.001 after 6 iterations.
In practice, the maximum number of iterations is set as 15 to guarantee a deviation of no more than $10^{-5}$.
Within each iteration, a convex optimization problem is solved with interior-point method, whose computation complexity is quite little.
Collecting the above results, the proposed algorithms can converge to the stationary point, i.e. local optimal value, with a low computation complexity.

\subsection{Performance analysis of the system}
In this subsection, the optimal bandwidth allocation and offloading ratio of the users are plotted versus the user index, to provide some insight on the optimal resource allocation strategy in the cooperative node assisted MEC system. 
Besides, the cost function in the system, i.e. the average power consumption to finish the users' tasks, is depicted as a function of system bandwidth $B$, computation capacity of the BS $f_B$ and task size collected in unit time $c_n$.
The results in this subsection present the performance of the system with various parameters, to provide insight on how to optimally offload the tasks (manage the communication resource) and allocate communication resource in the system.

\begin{table*}
	\centering
	\caption{Distance between the users and the cooperative node}
	\label{t:distance}
	\begin{tabular}{c|c|c|c|c|c}
		\hline
		\hline
		User No. & 1 & 2 & 3 & 4 & 5 \\
		\hline
		Distance (m) & 17.42 & 31.58 & 38.47 & 12.31 & 8.35 \\
		\hline
		\hline
		User No. & 6 & 7 & 8 & 9 & 10 \\
		\hline
		Distance (m) & 27.42 & 48.18 & 20.31 & 42.33 & 15.07 \\
		\hline
		\hline
	\end{tabular}
\end{table*}

\begin{figure}
	\begin{center}
		\includegraphics[angle=0,width=0.48 \textwidth]{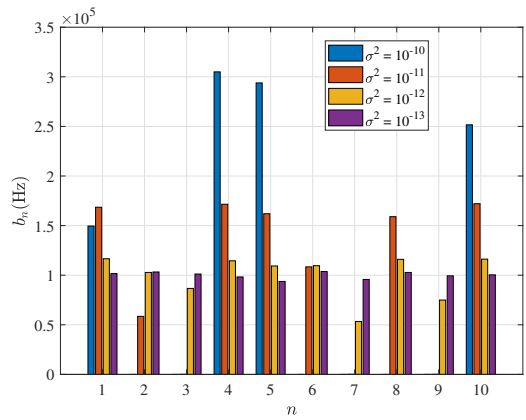}
	\end{center}
	\caption{Bandwidth allocation $b_n$ for different users in Case II with various settings of noise.}
	\label{f:bar_case_2_band}
\end{figure}
\begin{figure}
	\begin{center}
		\includegraphics[angle=0,width=0.48 \textwidth]{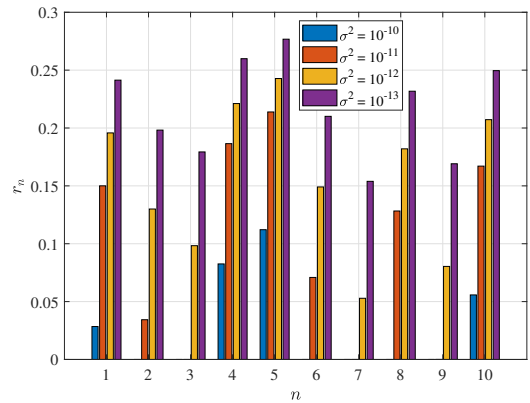}
	\end{center}
	\caption{Offloading ratio $r_n$ for different users in Case II with various settings of noise.}
	\label{f:bar_case_2_r}
\end{figure}

Looking into the scenario where there are 10 users assisted by a cooperative node.
To expose the impact of channel quality among different users on the bandwidth allocation and the task offloading ratio, 
the distance between the users and the cooperative node are randomly generated and presented with in Table \ref{t:distance}, while the effect of Rayleigh fading is temporarily left out.
In this way, the channel conditions between all users and the cooperative node can be is related to the distances alone.
In Fig. \ref{f:bar_case_2_band} and Fig. \ref{f:bar_case_2_r}, optimal solution of $b_n$ and $r_n$ with different power spectrum density of noise in Case II, i.e. the BS has finite computation capacity, are ploted respectively.
When the power spectrum density of noise is $10^{-10}$ mW/Hz, only user 1, user 4, user 5 and user 10 will offload the tasks, and they share the overall system bandwidth.
On the other hand, when the power spectrum density of noise is $10^{-13}$ mW/Hz, all the users can offload their task to the BS through the cooperative node.
In Fig. \ref{f:bar_case_2_band}, comparing the histograms of user pairs 1 v.s. 2 and 5 v.s. 6, some unintuitive result can be found:
In the situation where the channel noise power spectral density increases, users in closer proximity to the cooperative node can avail themselves of greater communication bandwidth. 
Conversely, when the channel noise power spectral density decreases, users situated farther from the cooperative node may be allocated more bandwidth than their counterparts in close proximity to the cooperative node.
The following interpretation can be drawn from the above results: 
In unfavorable channel conditions, communication resources are assigned to users with comparatively better channel conditions to enable them to reduce local energy consumption via task offloading.
In favorable channel conditions, more users tend to offload tasks to the BS to conserve energy. 
Therefore, the overall allocation of system bandwidth should be inclined towards users experiencing poorer channel conditions to ensure their success of offloading. 
Nevertheless, as depicted in Fig. \ref{f:bar_case_2_r}, users with superior channel conditions still tend to offload more tasks to the BS, 
confirming the fundamental observation in edge computing that users with better channel conditions are more prone to offloading tasks to the BS, which can result in more substantial energy savings.

\begin{figure}
	\begin{center}
		\includegraphics[angle=0,width=0.42 \textwidth]{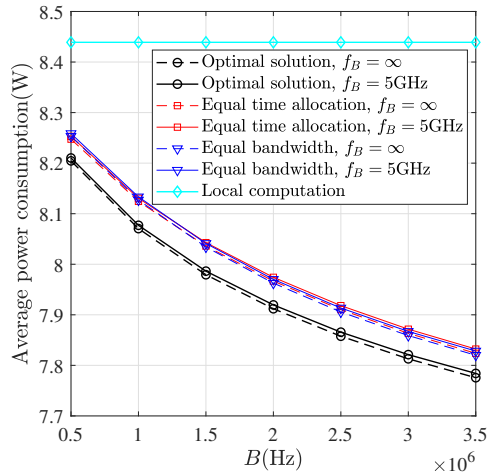}
	\end{center}
	\caption{Average power consumption versus system bandwidth $B$ in both Case I and Case II.}
	\label{f:figure_B}
\end{figure}
\begin{figure}
	\begin{center}
		\includegraphics[angle=0,width=0.42 \textwidth]{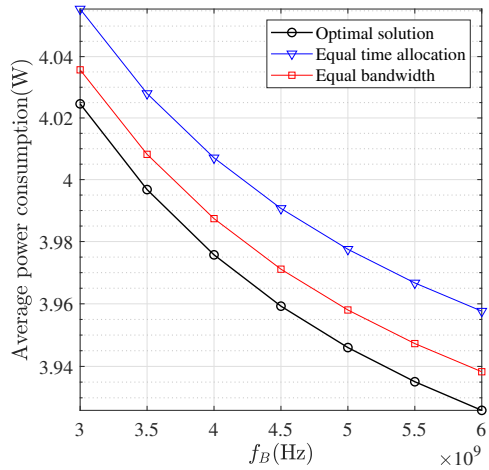}
	\end{center}
	\caption{Average power consumption versus the computation capacity of the BS $f_B$ in Case II.}
	\label{f:case_2_fB}
\end{figure}
\begin{figure}
	\begin{center}
		\includegraphics[angle=0,width=0.42 \textwidth]{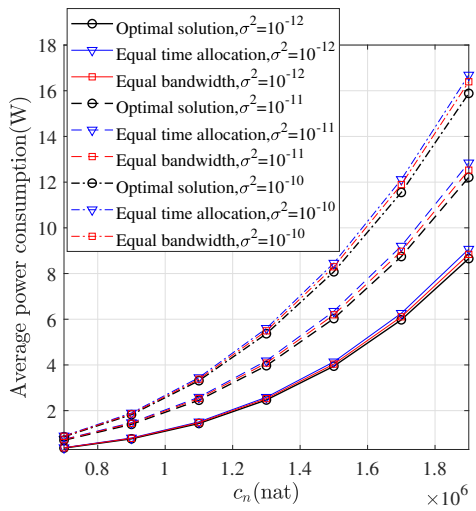}
	\end{center}
	\caption{Average power consumption versus data size generated in unit time $c_n$ Case II.}
	\label{f:case_2_c}
\end{figure}

In Fig. \ref{f:figure_B}, Fig. \ref{f:case_2_fB} and Fig. \ref{f:case_2_c}, 
considering that there are $N = 5$ users in the system and $\sigma = 10^{-10}$ mW/Hz, the average power consumption to finish the users' tasks (the cost function of the considered system), is depicted as a function of system bandwidth $B$, computation capacity of the BS $f_B$ and task size collected in unit time $c_n$.
To illustrate the optimality of the proposed method in this system, two benchmark schemes are considered:
1) $b_n = B/N, \forall n \in \mathcal{N}$. The users are allocated with equal bandwidth, which will be referred as 'Equal bandwidth' in the figures.
2) $t_2 = t_3$. The time consumption for transmission from the users to the cooperative node and from the cooperative node to the BS is consistent, which will be referred as 'Equal time allocation' in the figures.

By setting other parameters as default, Fig. \ref{f:figure_B} plots the average power consumption over the system bandwidth $B$, in which $f_B = \infty$ corresponds to Case I where the BS has abundant computation capacity and $f_B = 5 $ GHz is related to Case II where the BS has finite computation capacity.
Besides, the power consumption of offloading all the tasks to the BS and executing all the tasks locally is also investigated, in which the former one is too large to appear and the latter one is depicted in Fig. \ref{f:figure_B}.
Observing the curves in Fig. \ref{f:figure_B}, several results can be found:
\begin{enumerate}
	\item 
	When the system bandwidth increases, the average power consumption can be reduced, but the effect of reduction will shrink with increasing $B$. This is intuitive by referring to the expression of energy consumption in transmission: when the system bandwidth increases, the feasible region, either in Case I or Case II, is enlarged, thus the objective function can reach a lower value for certain.
	\item 
	The average power consumption does not show a great difference when the computation capacity of the BS varies from $f_B = 5 $ GHz to infinity.
	Given this observation, it can be inferred that the computation capacity of the BS is not a main restricting factors on the implement of MEC when the number of users is small, for the reason that the BS will finish the offloaded tasks with a negligible latency. In this case, adjusting the resource allocation by assuming that the BS has abundant capacity can be rational.
	On the other hand, for the situation where there are hundreds of users in the system whose tasks require extremely heavy computation burden, the result can be different.
\end{enumerate}

In Fig. \ref{f:case_2_fB} and Fig. \ref{f:case_2_c}, considering that the BS has a finite computation capacity, i.e. Case II, the average power consumption are ploted versus the BS 's computation capacity $f_B$ and the data size accumulated within unit time $c_n$, respectively.
When $f_B$ is increased, the feasible region of the optimization problem is relaxed and the average power consumption presents a reduction. 
However, this reduction is quite slight compared to that when increasing $B$, which can be explained by result 2) in the above.
When $c_n$ varies from $0.7 \times 10^6$ to $1.9 \times 10^6$, the average power consumption rises sharply by more than one order of magnitude whether the channel is good (the power spectrum density of noise is $10^{-12}$ mW/Hz) or bad (the power spectrum density of noise is $10^{-10}$ mW/Hz).
From this result, it can be observed that:
When the size of the task is already large, slight increase in task size will bring about an unacceptable increase in power consumption, even with the optimal solution in the system.

In summary, the results depicted in Figure \ref{f:figure_B}, Figure \ref{f:case_2_fB}, and Figure \ref{f:case_2_c} demonstrate that the proposed approach in this study can significantly reduce energy consumption compared to other benchmark schemes. 
Furthermore, comparing the alterations in the objective function exhibited in the aforementioned figures, it is apparent that although increasing resource investments such as communication bandwidth and computational power can temporarily mitigate the energy overhead associated with local computing, the present edge computing infrastructure will continue to confront great pressure to manage the complex task processing demands of various users.

%

\section{Conclusion} \label{s:conclusion}

\newpage

\vfill

\end{document}